\documentclass[twocolumn,secnumarabic,amssymb, amsmath, nofootinbib,tightenlines,
nobibnotes, aps, prl,epsfig]{revtex4}
\usepackage{graphicx}% Include figure files
\usepackage{dcolumn}% Align table columns on decimal point
\usepackage{bm}% bold math
\begin{document}
\preprint{APS/123-QED}
\title{Analytical approach for the approximate solution of the longitudinal structure function with respect to the GLR-MQ equation at Small $x$}% Force line breaks with \\

\author{B.Rezaei }
\altaffiliation{brezaei@razi.ac.ir}%Lines break automatically or can be forced with \\
\author{G.R.Boroun}%
 \email{grboroun@gmail.com; boroun@razi.ac.ir }
\affiliation{ Physics Department, Razi University, Kermanshah
67149, Iran}% \textbackslash\textbackslash
\date{\today}% It is always \today, today,

\begin{abstract}
We show that the nonlinear corrections to the longitudinal
structure function can be tamed the singularity behavior at low
$x$ values, with respect to GLR-MQ equations. This approach can
determined the shadowing longitudinal structure function based on
the shadowing corrections to the gluon and singlet quark structure
functions. Comparing our results with HERA data show that at very
low $x$ this behavior completely tamed by these corrections.
\end{abstract}
 %PACS, the Physics and Astronomy
                              %Classification Scheme.
 %Use showkeys class option if keyword
\maketitle
%%%%%%%%%%%%%%%%%%%%%%%%%%%%%%%%%%%%%%%%%%%%%%%%%%%%%%%%%%%%%%%%%

The measurement of the longitudinal structure function
$F_{L}(x,Q^{2})$ is of great theoretical importance, since it may
allow us to distinguish between different models describing the
QCD evolution at low-$x$. In deep- inelastic scattering (DIS), the
structure function measurements remain incomplete until the
longitudinal structure function $F_{L}$ is actually measured [1].
The longitudinal structure function in deep inelastic scattering
is one of the observable from which the gluon distribution can be
unfolded. At small $x$ values, the dominant contribution to
$F_{L}(x,Q^{2})$ comes from the gluon operators. Hence a
measurement of $F_{L}(x,Q^{2})$ can be used to extract the gluon
structure function and therefore the measurement of $F_{L}$
provides a
sensitive test of perturbative QCD [2].\\
In the region of moderate $x$ ($x{\geq}10^{-2}$) the well
established methods of operator expansion and renormalization
group equations have been applied successfully. The DGLAP
equations [3], which are based upon the sum of QCD ladder
diagrams, are the evolution equations in this kinematical region.
As observed, the longitudinal structure function can be related to
the gluon and sea- quark distribution. HERA shows [4-9] that the
gluon distribution function has a steep behavior in this region.
At small $x$, the problem is more complicated since recombination
processes between gluons in a dense system have to be taken into
account. As, the density of gluons and quarks becomes very high
and a new dynamical effect is expected to stop further growth of
the structure functions and so it has to be tamed by screening
effects. These screening effects are provided by multiple gluon
interaction which lead to nonlinear terms in the DGLAP equation.
These nonlinear terms reduce the growth of the gluon distribution
in this kinematic region where $\alpha_{s}$ is still small but the
density of partons becomes very large. Gribov, Levin, Ryskin,
Mueller and Qiu (GLR-MQ)[10,11] performed a detailed study of this
region. They argued that the physical processes of interaction and
recombination of partons become important in the parton cascade at
a large value of the parton density, and that these shadowing
corrections could be expressed in a new evolution equation (the
GLR-MQ equation). The main characteristic of this equation is that
it predicts a saturation of the gluon distribution at very small
$x$ [12,13]. This equation was based on two processes in a
parton cascade:\\
i)The emission induced by the QCD vertex $G{\rightarrow}G+G$ with
the probability which is proportional to $\alpha_{s}\rho$ where
$\rho(=\frac{xg(x,Q^{2})}{{\pi}R^{2}})$ is the density of gluon in
the transverse plane, ${\pi}R^{2}$ is the target area, and $R$ is the size of the target which the gluons populate;\\
ii)The annihilation of a gluon by the same vertex
$G+G{\rightarrow}G$ with the probability which is proportional to
$\alpha_{s}^{2}R^{2}\rho^{2}$, where $\alpha_{s}$ is probability
of the processes.\\
The main characteristic of this work is that it predicts a
saturation of the longitudinal structure function with respect to
the gluon and quark saturation distribution at very small $x$. The
starting point this work is the proof of the shadowing corrections
(SCs) to the parton behavior in QCD. The SCs for the
$F_{2}(x,Q^{2})$ proton structure function and $xg(x,Q^{2})$ gluon
distribution were computed considering in Ref.[14]. Here we
calculate this observable using the Altarelli- Marinelli equation
[15]. In perturbative QCD, the longitudinal structure function
$F_L(x,Q^2)$\ is proportional to $\alpha_s$. It is given to
leading order , by an integral over the quark and gluon
distributions:
\begin{eqnarray}
F^{sh}_L(x,Q^2)&=&\frac{\alpha_s(Q^2)}{\pi}(\frac{4}{3}\int_{x}^{1}\frac{dy}{y}(\frac{x}{y})^2F^{sh}_2(y,Q^2)\\\nonumber
&&+2\sum_{q}e_{q}^{2}\int_{x}^{1}\frac{dy}{y}(\frac{x}{y})^2(1-\frac{x}{y})G^{sh}(y,Q^2))
\end{eqnarray}
and $e_{q}$ are the quark charges.\\
As $x{\rightarrow}0$ the value of the gluon density becomes so
large that the annihilation of gluons becomes important. So, this
singular behavior is tamed by the shadowing effects. As, we assume
this behavior according to the Regge- like behavior:
\begin{equation}
f_{i}^{sh}(x,Q^{2})=A_{i}x^{-\lambda^{sh}},(i=g,q).
 \end{equation}
We note that at $x<x_{0}=10^{-2}$, shadowing and unshadowing
parton distribution behavior are equal. This picture allows us to
write GLR-MQ equations for the gluon and quark structure function
behavior at small $x$, as:
\begin{eqnarray}
\frac{dG^{sh}(x,Q^{2})}{dlnQ^{2}}=\frac{\alpha_{s}}{2\pi}\int^{1-\chi}_{0}{dz}G(\frac{x}{1-z},Q^{2})P_{gg}(\frac{1}{1-z})\nonumber\\
-\frac{81\alpha^{2}_{s}}{16R^{2}Q^{2}}\int^{1-\chi}_{0}\frac{dz}{1-z}[G(\frac{x}{1-z},Q^{2})]^{2},
\end{eqnarray}
and
\begin{eqnarray}
\frac{{\partial}F^{sh}_{2}(x,Q^{2})}{{\partial}lnQ^{2}}&=&\frac{{\partial}F_{2}(x,Q^{2})}{{\partial}lnQ^{2}}|_{DGLAP}\\\nonumber
&&-\frac{5}{18}\frac{27\alpha_{s}^{2}}{160R^{2}Q^{2}}[xg(x,Q^{2})]^{2}.
\end{eqnarray}
 Here $P_{gg}$ is
the gluon- gluon splitting function and $\chi=\frac{x}{x_{0}}$.
Where $x_{0}$ is the boundary condition that the gluon
distribution joints smoothly onto the unshadowed region. We
neglect the quark gluon emission diagrams due to their little
importance in the gluon rich low $x$ region and also the
nonsinglet contribution is negligible and can be ignored. To find
an analytical solution in leading order, we note that the
shadowing gluon  and quark distribution functions have the Regee-
like behavior corresponding to Eq.2. Inserting Eq.2 in Eqs.3 and
4, we can obtain the $Q^{2}$ shadowing logarithms slope of the
gluon  and quark distribution functions at low $x$, as:
\begin{eqnarray}
\frac{{\partial}G^{sh}(x,Q^{2})}{{\partial}lnQ^{2}}=\frac{3\alpha_{s}}{\pi}G^{s}\frac{1-\chi^{\lambda^{sh}}}{\lambda^{sh}}-\frac{81\alpha^{2}_{s}}{16R^{2}Q^{2}}(G^{s})^{2}\nonumber\\
\times\frac{1-\chi^{2\lambda^{sh}}}{2\lambda^{sh}},
\end{eqnarray}
and
\begin{eqnarray}
\frac{{\partial}F^{sh}_{2}(x,Q^{2})}{{\partial}lnQ^{2}}=\frac{5\alpha_{s}}{9\pi}T(\lambda_{g})G^{s}-\frac{5}{18}\frac{27\alpha_{s}^{2}}{160R^{2}Q^{2}}
[G^{s}]^{2},
\end{eqnarray}
where $
T(\lambda_{g})=\int_{\chi}^{1}z^{\lambda^{s}_{g}}[z^{2}+(1-z)^2]$.\\
Next, we combine terms and define
$\frac{{\partial}F^{sh}_{L}(x,Q^{2})}{{\partial}t}$ by:
\begin{eqnarray}
\frac{{\partial}F^{sh}_{L}(x,t)}{{\partial}t}&=&\frac{{d}\alpha_{s}}{{d}t}\frac{1}{\pi}(\frac{4}{3}\int_{x}^{1}\frac{dy}{y}(\frac{x}{y})^2F^{sh}_2(y,t)\\\nonumber
&&+2\sum_{q}e_{q}^{2}\int_{\chi}^{1}\frac{dy}{y}(\frac{x}{y})^2(1-\frac{x}{y})G^{sh}(y,t))\\\nonumber
&&+\frac{\alpha_s(t)}{\pi}(\frac{4}{3}\int_{\chi}^{1}\frac{dy}{y}(\frac{x}{y})^2\frac{{\partial}F^{sh}_2(y,t)}{{\partial}t}\\\nonumber
&&+2\sum_{q}e_{q}^{2}\int_{\chi}^{1}\frac{dy}{y}(\frac{x}{y})^2(1-\frac{x}{y})\frac{{\partial}G^{sh}(y,t)}{{\partial}t}),
\end{eqnarray}
where
$\alpha^{LO}_{s}(Q^{2})=\frac{4\pi}{\beta_{0}\ln(\frac{Q^{2}}{\Lambda^{2}})}$,
$\beta_{0}=\frac{1}{3}(33-2N_{f})$ and $N_{f}$ being the number of
active quark flavors ($N_{f}=4$), also
t=ln$(\frac{Q^{2}}{\Lambda^{2}})$ (that $\Lambda$ is the QCD cut-
off parameter).\\
To obtain a differential equation for $F^{sh}_{L}(x,t)$, we
writing out the sum and the splitting function explicitly, we find
an inhomogeneous first- order differential equation which
determine $F^{sh}_{L}(x,t)$ in terms of $G^{sh}(x,t)$. Eq.(7) can
be put in the following form, as:
\begin{eqnarray}
\frac{{\partial}F^{sh}_{L}(x,t)}{{\partial}t}=\frac{dLn\alpha_{s}}{dt}F^{sh}_{L}(x,t)+F^{sh}_{0L}(x,t),
\end{eqnarray}
where, to simplify the notation in Eq.7, we define
$F^{sh}_{0L}(x,t)$ by:
\begin{eqnarray}
F^{sh}_{0L}(x,t)=\frac{\alpha_s(t)}{\pi}\frac{4}{3}\int_{\chi}^{1}\frac{dy}{y}(\frac{x}{y})^2\frac{{\partial}F^{sh}_{2}(y,Q^{2})}{{\partial}lnQ^{2}}[Eq.6]\\\nonumber
+\frac{\alpha_s(t)}{\pi}\frac{20}{9}\int_{\chi}^{1}\frac{dy}{y}(\frac{x}{y})^2(1-\frac{x}{y})\frac{{\partial}G^{sh}(y,Q^{2})}{{\partial}lnQ^{2}}[Eq.5].
\end{eqnarray}
Eq.8 has the advantage that it determines the shadowing
longitudinal structure function completely in terms of hot- spot
point. In [14] this procedure was applied to obtain the shadowing
corrections (SCs)
to the nucleon gluon distribution and the singlet structure function.\\
We have found that we can parametrize the shadowing longitudinal
structure function calculated using the shadowing gluon
distribution function  and shadowing singlet quark structure
function determined as a linear equation into the running coupling
constant whose tamed as $x{\rightarrow}0$. Indeed Eq.(8) is of the
form:
\begin{eqnarray}
\frac{{\partial}F^{sh}_{L}(x,t)}{{\partial}t}+\eta(t)F^{sh}_{L}(x,t)=F^{sh}_{0L}(x,t),
\end{eqnarray}
where $\eta(t)=-\frac{dLn\alpha_{s}(t)}{dt}$, then we have a
linear equation of the integrating factor type. The factor
$R(t)=exp[\int\eta(t)dt]$ can be calculated and it is easily shown
that:
\begin{eqnarray}
\frac{{\partial}(F^{sh}_{L}(x,t)R(t))}{{\partial}t}=F^{sh}_{0L}(x,t)R(t).
\end{eqnarray}
So that it is straightforward to integrate both sides with respect
to $t$, we find that:
\begin{eqnarray}
F^{sh}_{L}(x,t)=\alpha_{s}(t){\int}F^{sh}_{0L}(x,t)/\alpha_{s}(t)dt.
\end{eqnarray}
This equation shows the dependence of the shadowing longitudinal
structure function on the strong constant coupling and the
shadowing gluon distribution function. Therefore, this equation is
an attempt to include the full expression of the SCs. At small $x$
the nonlinear term with the shadowing gluon distribution function
is the dominant one. Consequently, the expression (12) can be
tamed reasonably by the shadowing corrections. This equation
demonstrates the close relation between the shadowing longitudinal
structure function and the shadowing gluon distribution function
at hot spot point. Therefore, we expect the shadowing longitudinal
structure function to be sensitive to the shadowing corrections in
the HERA kinematic region. We emphasize at this point that the
shadowing gluon distribution function $G^{sh}(x,t)$ is completely
determined based on $\lambda^{sh}_{g}$ and its derivative with
respect to t through the expressions in Eq.10 at Ref.14. The
results are shown in Fig.1 for $Q^{2}=20\hspace{0.1cm}GeV^{2}$ at
hot-spot point $R=2\hspace{0.1cm}GeV^{-1}$. The simple
conclusions, which could be obtained from the present plot, is the
following. Our results at hot- spot for
$Q^{2}=20\hspace{0.1cm}GeV^{2}$ give
 values comparable of the shadowing longitudinal structure function  that are comparing with
experimental data and QCD analysis fits. This grow  with the
rapidity $1/x$. Our data show that shadowing  longitudinal
structure function increase as x decreases, that its corresponding
with PQCD fit at low x, but this behavior tamed with respect to
nonlinear terms at GLR-MQ equation.\\
Our conclusion is that the longitudinal structure function $F_{L}$
is a good observable to isolate the shadowing corrections at HERA,
since new effects must occur in low $x$ kinematical region. In
this letter we estimate the shadowing corrections to the
longitudinal structure function into shadowing correction to the
gluon and singlet distribution functions with respect to GLR-MQ
equations. As the behavior of this observable is directly
dependence on the behavior of the gluon and singlet distribution
functions and, therefore, strongly sensitive to the shadowing
corrections. We show that the obtained results for the shadowing
longitudinal structure function at small-$x$ strongly modified by
shadowing corrections as this growth tamed by the shadowing
effects.\\

%%%%%%%%%%%%%%%%%%%%%%%%%%%%%%%%%%%%%%%%%%%%%%%%%%%%%%%%%%%%%%%%%%%%%%%%
\textbf{References}\\
\hspace{2cm}1. A.Gonzalez-Arroyo, C.Lopez, and F.J.Yndurain, phys.lett.B\textbf{98}, 218(1981).\\
\hspace{2cm}2. A.M.Cooper- Sarkar, G.Inglman, K.R.Long, R.G.Roberts, and D.H.Saxon , Z.Phys.C\textbf{39}, 281(1988).\\
  R.G.Roberts, The structure of the proton, (Cambridge University Press 1990)Cambridge.\\
  \hspace{2cm}3. Yu.L.Dokshitzer, Sov.Phys.JETP {\textbf{46}},
641(1977); G.Altarelli and G.Parisi, Nucl.Phys.B \textbf{126},
298(1977); V.N.Gribov and L.N.Lipatov,
Sov.J.Nucl.Phys. \textbf{15}, 438(1972).\\
  \hspace{2cm}4. S.Aid et.al, $H1$ collab. phys.Lett. {\bf B393}, 452-464 (1997).\\
\hspace{2cm}5. R.S.Thorne, phys.Lett. {\bf B418}, 371(1998).\\
\hspace{2cm}6. C.Adloff et.al, $H{1}$ Collab., Eur.Phys.J.C\textbf{21}, 33(2001).\\
\hspace{2cm}7. N.Gogitidze et.al, $H{1}$ Collab., J.Phys.G\textbf{28}, 751(2002).\\
\hspace{2cm}8. A.V.Kotikov and G.Parente, JHEP \textbf{85},
17(1997);
Mod.Phys.Lett.A\textbf{12}, 963(1997).\\
\hspace{2cm}9. C.Adloff,$H1$ collab. phys.Lett. {\bf B393}, 452(1997).\\
10. L.V.Gribov, E.M.Levin and M.G.Ryskin, Phys.Rep.\textbf{100},
 1(1983).\\
11.A.H.Mueller, J.Qiu, Nucl.Phys.{\bf B268}, 427(1986).\\
12. A.L.A.Filho. M.B.Gay Ducati and V.P.Goncalves,
 Phys.Rev.D\textbf{59},054010(1999).\\
13.K.J.Eskola et al., Nucl.Phys.B\textbf{660},211(2003).\\
14.G.R.Boroun , Eur. Phys.J.A\textbf{42}, 251(2009); Eur. Phys.J.A\textbf{43}, 335(2010).\\
 15. G.Altarelli and G. Martinelli, Phys.Lett.B\textbf{76},89(1978).\\
 16. A.Vogt, S.Moch, J.A.M.Vermaseren, Nucl.Phys.B \textbf{691},
129(2004).\\
17. S.Moch, J.A.M.Vermaseren, A.vogt, Phys.Lett.B \textbf{606},
123(2005).\\
18. A.D.Martin, R.G.Roberts, W.J.Stirling,R.Thorne, Phys.Lett.B
\textbf{531}, 216(2001).\\
19. E.L.Berger, M.M.Block and C.I Tan, Phys.Rev.Lett\textbf{98},
242001(2007).\\
20. M.M.Blok, L.Durand and D.W.McKay, Phys.Rev.D\textbf{77},
094003(2008).\\
21. A.Donnachie and P.V.Landshoff, Phys.Lett.B {\bf550}, 160(2002 ); P.V.Landshoff,hep-ph/0203084.\\
%%%%%%%%%%%%%%%%%%%%%%%%%%%%%%%%%%%%%%%%%%%%%%%%%%%%%%%%%%%%%%%

%%%%%%%%%%%%%%%%%%%%%%%%%%%%%%%%%%%%%%%%%%%%%%%%%%%%%%
\subsection{Figure captions }
Fig 1:The values of the shadowing longitudinal structure function
at $Q^{2}=20\hspace{0.1cm} GeV^{2}$ with
    $R=2\hspace{0.1cm}GeV^{-1}$ (square) that accompanied with model error by solving the
GLR-MQ evolution equation
  that compared with H1 Collab. data
    (up and down triangle). The error on the  H1
 data is the total uncertainty of the determination of
 $F_{L}$ representing the statistical, the systematic and the model errors added in quadrature.
Circle data are the MVV prediction [16-18
]. The solid line is the
NLO QCD fit to the H1 data for $y<0.35$ and
  $Q^{2}{\geq}3.5\hspace{0.1cm}GeV^{2}$.  The dot line is the DL fit [21] and the dash line is a
   QCD fit with respect to LO gluon distribution function from Berger [19-20] fit. \\
\begin{figure}
\includegraphics[width=1\textwidth]{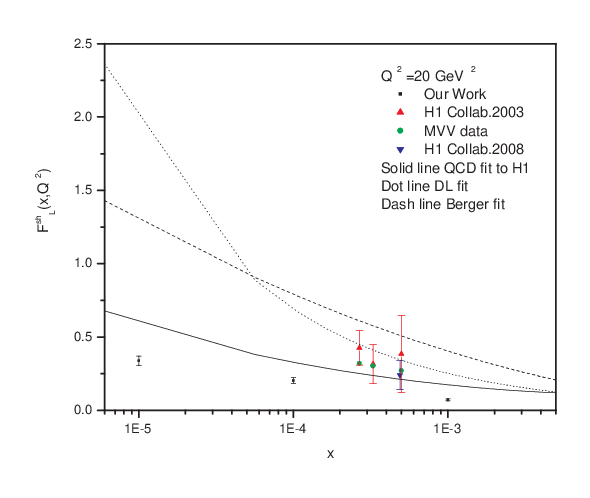}
\caption{} \label{Fig1}
\end{figure}

%%%%%%%%%%%%%%%%%%%%%%%%%%%%%%%%%%%%%%%%%%%%%%%%%%%%%%%%%%%%
\end{document}